\documentclass[twocolumn]{webofc}
\usepackage[varg]{txfonts}   
\usepackage{aas_macro}
\usepackage{mathrsfs}

\newcommand{\msun}{\rm M_{\odot}}
\begin{document}
\title{High-energy emissions from neutron star mergers}
%
%

\author{\firstname{Shigeo S.} \lastname{Kimura}\inst{1,2,3}\fnsep\thanks{\email{szk323@psu.edu}}
}

\institute{Department of Physics, Pennsylvania State University, University Park, Pennsylvania, 16802, USA
\and
    Center for Particle and Gravitational Astrophysics, Pennsylvania State University, University Park, Pennsylvania, 16802, USA
\and
Department of Astronomy \& Astrophysics, Pennsylvania State University, University Park, Pennsylvania, 16802, USA\\
          }

\abstract{%
In 2017, LIGO-Virgo collaborations reported detection of the first neutron star merger event, GW170817, which is accompanied by electromagnetic counterparts from radio to gamma rays. Although high-energy neutrinos were not detected from this event, mergers of neutron stars are expected to produce such high-energy particles. Relativistic jets are launched when neutron stars merge. If the jets contain protons, they can emit high-energy neutrinos through photomeson production. In addition, neutron star mergers produce massive and fast ejecta, which can be a source of Galactic high-energy cosmic rays above the knee. We briefly review what we learned from the multi-messenger event, GW170817, and discuss prospects for multi-messenger detections and hadronic cosmic-ray production related to the neutron star mergers.
}
\maketitle
\section{Introduction}\label{sec:intro}

Binary neutron star (BNS) mergers have been actively discussed as sources of multi-messenger astrophysics for a long time. A close BNS can merge within the Hubble time because the orbit of the BNS decreases due to emission of gravitational waves (GWs). This is confirmed by the observation of the binary pulsar \cite{Hulse:1974eb}. 

When the BNS merges, a compact object, either a black hole or massive neutron star, is left. The remnant object accrets surrounding material and releases a large amount of gravitational energy, which is expected to launch relativistic jets. If the jets accelerate electrons by dissipating their kinetic energy, they emit gamma-rays, which can be observed as a short gamma-ray burst (SGRB) \cite{ELP89a,Nak07a,Ber14a}.  
The r-process elements are also expected to be produced by the neutron star mergers because the neutron stars consist of neutron-rich material \cite{LS74a}. According to the numerical relativity simulation, the mergers create massive ejecta \cite{HKK13a}, and decay of the radioactive nuclei powers the optical/infrared transients (kilonova/macronova; hereafter macronova) \cite{LP98a,TH13a,BK13a,Met16a}. Optical and infrared observations of SGRB afterglows give some hints of macronovae (e.g., \cite{HKT13a}). 
The jets and ejecta of neutron star mergers form forward shocks through interaction with ambient matter. Electrons are accelerated at the shocks, which produce a broad band afterglow emission through the synchrotron emission \cite{MR97a,NP11a,HNH16a}. 
The BNS mergers can be sources of hadronic high-energy particles. If the jets of SGRBs contain protons, they are accompanied with high-energy neutrinos \cite{WB97a}. The kinetic energy of the jets dissipates in the dissipation region through some plasma processes, such as shocks \cite{RM94a} or magnetic reconnection \cite{MU12a}. The protons are accelerated to very high energy there and interact with the target photons observed as SGRBs, producing pions that decay to the neutrinos. Also, the ejecta of macronovae can accelerate cosmic-ray protons beyond the knee, because they are faster than supernova ejecta which accelerate the Galactic cosmic rays below the knee.

The multi-messenger observation of GW170817  confirmed most of the pictures above \cite{LIGO17d}. This event was detected by the GWs, radio waves, optical, ultraviolet (UV), infrared (IR), X-rays, and MeV gamma-rays. However, GeV and TeV gamma-rays and neutrinos are not detected, despite that the BNS mergers are expected to emit these high-energy particles. The GW170817 observations give us the physical quantities of the macronova ejecta, which enables us to discuss the hadronic high-energy processes related to BNS mergers in more quantitative manner. In this paper, we briefly review what we learned from GW170817, and discuss future prospects for high-energy neutrino detections and hadronic cosmic-ray production from neutron star mergers. The main topics of this paper (Section \ref{sec:Neutrinos} and \ref{sec:NSMRs}) are published in \cite{KMM17b,KMB18a,KMM18a}.

\section{GW170817}\label{sec:GW170817}

In 2017, LIGO-Virgo collaborations reported multi-messenger detection of a BNS merger event (GW170817; \cite{LIGO17d}). First, LIGO-Virgo detected GW signals from a inspiraling binary of total mass of 2.8 $\msun$ and mass ratio of $\ge0.7$. This tells us that the binary consists of two neutron stars \cite{LIGO17c}. From the GW data analysis, the luminosity distance is estimated to be 40 Mpc. The localization of the event is around 30 degree$^2$, which is much smaller than the previous binary black hole merger events owing to the observation by three detectors.

1.7 sec after the merger, the {\it Fermi} Gamma-ray Burst Monitor (GBM) and the SPectrometer on board {\it INTEGRAL} Anti-Coincidence Shield (SPI-ACS) detected a short gamma-ray burst, GRB 170817A \cite{LIGO17e}. This prompt gamma-ray emission confirmed the BNS merger paradigm of SGRBs. We learned that at least some fraction of SGRBs occur through  BNS mergers. However, the luminosity of GRB 170817A is $L_{\rm iso}\simeq 1.6\times10^{46}\rm~erg~s^{-1}$, which is much lower than the typical SGRBs occurred at cosmological distance, $L_{\rm iso}\sim10^{51}\rm~erg~s^{-1}$. The emission mechanism of such low-luminosity prompt gamma-rays is still controversial. One possibility is that GW 170817 is an off-axis event of the classical SGRB (e.g., \cite{IN18a,LK18a}). Another is the shock breakout from the macronova ejecta (e.g., \cite{KNS17a,GNP18a}). 

Eleven hours later, the optical counterpart is identified at NGC 4993, a galaxy located around 40 Mpc away from the Earth \cite{CFK17a,MASTER17a,DES17a,DLT4017a}. The distance is consistent with that obtained from the GW analysis. Initially, the transient evolves rapidly in both luminosity and color, compared to supernovae. The color becomes redder in later time (e.g. \cite{JGEM17a}). This rapid and redward evolutions are consistent with the theoretical modeling of a macronova powered by radioactivity of r-process elements. The modeling tells us that two-component models fit the data well: the fast-light component with a low opacity ($M\sim 0.01\msun-0.03\msun$, $V\sim 0.3c$, $\kappa\sim1\rm~cm^2~g^{-1}$) and the slow-heavy component with a high opacity ($M\sim0.03\msun-0.05\msun$, $V\sim 0.1c-0.2c$, $\kappa\sim10\rm~cm^2~g^{-1}$) (e.g., \cite{DPS17Sa,MRK17a,JGEM17b,SFH17a,KMB17a}). To produce the massive, fast, and low-opacity ejecta, the remnant central object should be a temporal hypermassive neutron star. If the hypermassive neutron star exists for a long time, the strong X-ray and gamma-rays should be detected weeks to months after the merger \cite{MTF18a}. Since we did not detect such signals, the hypermassive neutron star should collapse to a black hole after the macronova ejecta is produced.

The X-ray and radio counterparts are detected 9 days and 16 days after the merger, respectively \cite{TPV17a,MBF17a,HNR17a,HCM17a,ABF17a,ALOP17a}. The spectrum of the afterglow is consistent with the synchrotron emission from electrons with a single power-law distribution of the index $\simeq 2.2$. The light curve shows the gradual increase in both radio and X-ray bands, $L\propto t^{0.7}$ for more than 100 days after the merger (e.g. \cite{MNH18a,MAX18a}). This ruled out the classical top-hat jet model seen from off-axis, and a radial or azimuthal structure is demanded. To explain this feature, two possibilities are intensely discussed: the quasi-spherical cocoon (or ejecta) with a radial structure \cite{MNH18a,HKS18a} and the azimuthally structured jet \cite{MAX18a,LPM18a}. Resolving the emission region is a smoking gun for distinguishing these two models. 230 days after the merger, the VLBI radio observation detected a superluminal motion of the emission region, which undoubtedly indicates that the emission region has a relativistic velocity \cite{MDG18a}. Also, they found that the emission region is so compact that even the VLBI observations cannot  resolve it. These results favor the structured jet model rather than the quasi-spherical cocoon model. This supports that GW170817 is a canonical SGRB seen from off-axis. The light curve of the afterglow started to fade around 150 days with a decreasing rate of $\sim t^{-2.2}$ \cite{MFD18a}. Such a rapid decreasing also favors the structured jet model, because the quasi-spherical cocoon models predict slower decreasing rates.

The higher-energy gamma rays and neutrinos from GW170817 are not detected, despite the intense search by several collaborations \cite{HESS17a,Fermi18a,IceCube17c,SK18a}. We discuss the neutrino emissions and implications of this result in the next section.


\section{High-energy neutrinos from neutron star mergers}\label{sec:Neutrinos}

From GW170817, we learned that BNS mergers indeed produce relativistic jets. These jets are expected to produce high-energy neutrinos if they contain protons. Since the emission is strongly beamed toward the jet direction, detection from off-axis events are very challenging. Here, we discuss the detectability of the neutrinos from on-axis events.

After BNS mergers, the jets launched from the central engine interact with the ejecta of macronovae, and the outcomes can be classified into two cases: the successful SGRBs with late-time activities (Section \ref{sec:SGRBs}) and the failed SGRBs with choked jets (Section \ref{sec:trans}). For the successful jet case, the jets successfully penetrate the ejecta of macronovae, resulting in the canonical SGRBs.  For the failed SGRBs case, the jets fail to penetrate the ejecta, and bright gamma-rays are not detected from this system. Only the neutrinos are able to come out from the inside of the ejecta. We call these neutrinos ``trans-ejecta neutrinos''. See \cite{KMM17b} and \cite{KMB18a} for details of the neutrinos from the successful jet and the choked jet cases, respectively.

We calculate the neutrino fluence using phenomenological formula. We consider a single power-law proton spectrum with a spectral index $s=2$: 
\begin{equation}
 E_p^2\frac{dN_p^{\rm iso}}{dE_p}\approx \frac{\xi_{\rm acc}\mathscr{E}^{\rm iso}_{\rm rad}}{\ln(E_{p,\rm max}/E_{p,\rm min})} \exp\left(-{E_p\over E_{p,\rm max}}\right),\label{eq:Ep2dNdEp}
\end{equation}
where $E_p$ is the proton energy at observer frame, $\xi_{\rm acc}$ is the baryon loading factor, $\mathcal E^{\rm iso}_{\rm rad}$ is the isotropic equivalent gamma-ray energy fluence, and $E_{p,\rm max}$ and $E_{p,\rm min}$ are the maximum and minimum proton energy, respectively. $E_{p,\rm max}$ is given by the balance between cooling and acceleration. The muon neutrino spectrum produced by pion decay is  written as 
\begin{equation}
 E_{\nu_\mu}^2\frac{dN_{\nu_\mu}}{dE_{\nu_\mu}}\approx\frac18f_{p\gamma}f_{\rm sup\pi}{E_p^2}\frac{dN_p}{dE_p},\label{eq:nu_mu}
\end{equation}
where $E_{\nu_\mu}\simeq 0.05E_p$ is the muon neutrino energy at the observer frame, $f_{p\gamma}$ is the effective optical depth for the photomeson production, and $f_{\rm sup\pi}$ is the suppression factor by pion coolings. The electron neutrino and muon antineutrino spectra produced by muon decay are given by 
\begin{equation}
 E_{\nu_e}^2\frac{dN_{\nu_e}}{dE_{\nu_e}}\approx E_{\overline\nu_\mu}^2\frac{dN_{\overline\nu_\mu}}{dE_{\overline\nu_\mu}}\approx\frac{1}{8}f_{p\gamma}f_{\rm sup\pi}f_{\rm sup\mu}E_p^2\frac{dN_p}{dE_p},\label{eq:nu_e} 
\end{equation}
where $E_{\nu_e}\simeq 0.05E_p$ and $E_{\overline\nu_\mu}\simeq0.05E_p$ are the electron neutrino and muon antineutrino energies, respectively,  $f_{\rm sup\mu}$ is the suppression factor by muon coolings. We appropriately take into account the energy-dependent cross-section of photomeson production \cite{MN06b}, proton cooling processes (synchrotron, adiabatic cooling, and photomeson production for the successful and choked jets, Bethe-Heitler process for the choked jets), pion cooling processes (synchrotron and adiabatic cooling for the successful and choked jets, proton-pion inelastic collisions for the choked jets), and muon cooling processes (synchrotron and adiabatic cooling for the successful and choked jets). During the propagation from the source to the Earth, the neutrino oscillation changes the flavor ratio, which is calculated using the tri-bimaximal mixing matrix (e.g., \cite{HPS02a})
\begin{equation}
 \phi_{\nu_e+\overline\nu_e}=\frac{10}{18}\phi_{\nu_e+\overline\nu_e}^0+\frac{4}{18}(\phi_{\nu_\mu+\overline\nu_\mu}^0+\phi_{\nu_\tau+\overline\nu_\tau}^0),
\end{equation}
\begin{equation}
 \phi_{\nu_\mu+\overline\nu_\mu}=\frac{4}{18}\phi_{\nu_e+\overline\nu_e}^0+\frac{7}{18}(\phi_{\nu_\mu+\overline\nu_\mu}^0+\phi_{\nu_\tau+\overline\nu_\tau}^0),
\end{equation}
where $\phi_i^0=(dN_i/dE_i) / (4\pi{d_L^2})$ is the neutrino fluence without the oscillation and $d_L$ is the luminosity distance.

We discuss the prospects for the neutrino detection coincident with GWs by IceCube and IceCube-Gen2. The expected number of neutrino detection is estimated to be 
\begin{equation}
 \overline{\mathcal{N}_\mu}=\int\phi_\nu{A}_{\rm{eff}}(\delta,~E_\nu)dE_\nu,
\end{equation}
where $A_{\rm eff}$ is the effective area and $\delta$ is the declination angle.The effective area for $\nu_\mu$-induced events with IceCube is given by \cite{IceCube17b}. For lower energy of $\lesssim1$ PeV, the effective area for up-going + horizontal events is larger than that for down-going events, because the atmospheric muons are shielded by the Earth. For the higher energies, the neutrinos are also blocked by the Earth, so the effective area for the down-going events is larger. For IceCube-Gen2, we use $10^{2/3}$ larger effective area than that for IceCube. Although such a simple scaling might not be a good approximation for the down-going events, this treatment suffices for a demonstration purpose. The detection probability of $k$ neutrino events is given by the Poisson distribution.

\subsection{Neutrinos from SGRBs}\label{sec:SGRBs}

The observed SGRBs are followed by afterglows. The classical afterglow theory based on the forward shock model predicts a decreasing light curve with a single power-law function \cite{SPN98a,Mes06a,KZ15a}. SGRBs are often followed by the afterglow with flat light curves (extended emission for $t\sim10^2-10^3$ sec, and plateau emission for $t\sim 10^3-10^4$ sec \cite{NB06a,Swift11a,KIS17a}). X-ray flares are also observed during the afterglow of SGRBs \cite{MCG11a}. Since the classical forward shock models have difficulty to explain the time variability of these emissions, they are likely to originate from prolonged central engine activities \cite{IKZ05a}. The energy fluence of the late-time emissions are comparable to that of the prompt emission \cite{KIS17a}, so they are important components of this system with regard to energetics.

\begin{figure}
\centering
\includegraphics[width=\linewidth]{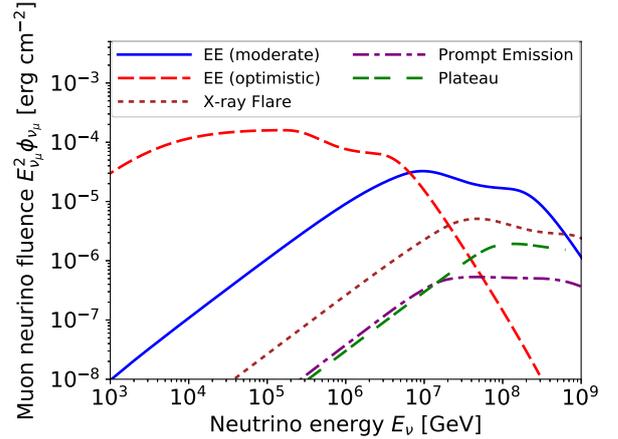}
\caption{Neutrino fluences from extended emissions (EEs), plateau emission, prompt emission, and X-ray flare of SGRBs at the luminosity distance $d_L=300$ Mpc. We consider optimistic and moderate cases for EEs. This figure is reproduced from \cite{KMM17b}.}
\label{fig:sgrb-fluence}       
\end{figure}

The late time emissions are also expected to produce high-energy neutrinos. Since these components can have a lower Lorentz factor and a lower break energy than those for the prompt emission, the photon density can be higher, leading to the efficient neutrino production.  We consider broken power-law spectra for the target photons with indices $\alpha=0.5$ and $\beta=2.0$ below and above the break energy, respectively. We calculate the neutrino spectra from the prompt emission, the extended emissions (two cases), the plateau emission, and the X-ray flare with the model parameters tabulated in Table \ref{tab:sgrb-param}. These parameters are obtained from the observations, although some of them are not constrained very well. The resulting physical quantities are tabulated in Table \ref{tab:sgrb-quant}. The late-time emissions can accelerate protons up to several EeVs to 100 EeV, depending on the component.  Figure \ref{fig:sgrb-fluence} shows the neutrino spectra with the baryon loading factor $\xi_{\rm acc}=10$. We can see two breaks in the spectra. The first break appears due to the change of the photon spectral index. The second break comes from the pion synchrotron cooling.  Owing to the higher luminosity and the lower Lorentz factor, the extended emissions have denser photon fields. Therefore, the extended emission is the most efficient neutrino emitter of the four. 

\begin{table*}
\centering
\caption{Used parameters for each component of SGRBs. $\Gamma$ is the Lorentz factor of the jet, $L_{\gamma,\rm iso}^*$ and $\mathscr E_{\gamma,\rm iso}^*$ are the isotropic equivalent luminosity and energy fluence in the observed energy band, $r_{\rm diss}$ is the dissipation radius, $ E_{\gamma,\rm pk}$ is the observed break energy of the photon spectrum, and the last column shows the energy band of the SGRB observations. This table is adapted from \cite{KMM17b}.   }
\label{tab:sgrb-param}       
\begin{tabular}{lllllll}
\hline
parameters & $\Gamma$ &$L_{\gamma,\rm iso}^{*}$ [$\rm~erg~s^{-1}$] & $\mathscr{E}_{\gamma,\rm iso}^{*}$ [erg] & $r_{\rm diss}$ [cm] & $E_{\gamma,\rm pk}$ [keV] & energy band [keV] \\
\hline
EE-mod & 30 & 3$\times10^{48}$ & $10^{51}$ & $10^{14}$ & 1 & 0.3--10 \\
EE-opt & 10 & 3$\times10^{48}$ & $10^{51}$ & 3$\times10^{13}$ & 10 & 0.3--10 \\
prompt & $10^3$ & $10^{51}$ & $10^{51}$ &  3$\times10^{13}$ & 500 & 10--$10^3$ \\
flare & 30 & $10^{48}$ &  3$\times10^{50}$ &  3$\times10^{14}$ & 0.3 & 0.3--10 \\
plateau & 30 & $10^{47}$ &  3$\times10^{50}$ &  3$\times10^{14}$ & 0.1 & 0.3--10 \\
\hline
\end{tabular}
\end{table*}

\begin{table*}
\centering
\caption{Resulting physical quantities for each component of SGRBs. $B$ is the magnetic field in the dissipation region, $L_{\gamma,\rm iso}$ and $\mathscr{E}_{\gamma,\rm iso}$ are the total isotropic equivalent luminosity and energy fluence, $E_{p,M}$ is the maximum energy for protons, $E_{\nu,\rm\mu}$ is the critical neutrino energy above which the muon cooling is effective, $E_{\nu,\rm\pi}$ is the break energy of neutrino spectrum due to pion cooling. This table is adapted from \cite{KMM17b}.   }
\label{tab:sgrb-quant}       
\begin{tabular}{lllllll}
\hline
 quantities& ${B}$ [G] &$L_{\gamma,\rm iso}$ [$\rm~erg~s^{-1}$] &$\mathscr{E}_{\gamma,\rm iso}$ [erg] &$E_{p,M}$ [EeV] &$E_{\nu,\rm\mu}$ [EeV] &$E_{\nu,\rm\pi}$ [EeV] \\
\hline
EE-mod & 2.9$\times10^{3}$ & 1.2$\times10^{49}$ & 3.8$\times10^{51}$ & 21 & 0.020 & 0.28\\
EE-opt & 5.0$\times10^{4}$ & 3.4$\times10^{49}$ & 1.1$\times10^{52}$ & 6.0 & $3.9\times10^{-4}$ & 5.4$\times10^{-3}$\\
prompt & 6.7$\times10^{3}$ & 6.1$\times10^{51}$ & 6.1$\times10^{51}$ & 60 & 0.29 & 4.0 \\
flare & 5.3$\times10^{2}$ & 3.5$\times10^{48}$ & 1.0$\times10^{51}$ & 25 & 0.11 & 1.5\\
plateau & 1.8$\times10^{2}$ &  3.8$\times10^{47}$ & 1.1$\times10^{51}$ & 13& 0.33 & 4.6\\
\hline
\end{tabular}
\end{table*}

We discuss the neutrino detection probability coincident with GWs. Hereafter, we focus on the neutrinos from the extended emissions. Since the neutrino fluences strongly depend on the Lorentz factor, we consider the distribution of the Lorentz factor, assuming that the log-normal function describes the distribution function:
 \begin{equation}
F(\Gamma)=\frac{dN_\Gamma}{d\ln\Gamma}=F_0\exp\left(-\frac{(\ln(\Gamma/\Gamma_0))^2}{2(\ln(\sigma_\Gamma))^2}\right),\label{eq:jetdist}
\end{equation}
where $F_0$ is the normalization factor, $\Gamma_0$ is the mean Lorentz factor, and $\sigma_\Gamma$ is the dispersion of the distribution. We fix the other parameters to those of EE-mod and EE-opt shown in Table \ref{tab:sgrb-param}. We consider the merger events at 300 Mpc, which is the detection horizon of the design sensitivity for advanced LIGO. Using the local event rate obtained by the SGRB observations, $4-10\rm~Gpc^{-3}~yr^{-1}$ \cite{NGF06a,WP15a}, we expect 2-5 extended emissions within 10 years if half of SGRBs are followed by extended emissions (cf. \cite{2015ApJ...805...89L}). Assuming that all the merger events at 300 Mpc are detected by GWs, we tabulate the resulting probabilities of neutrino detection coincident with GWs  within 10 years of operation in Table \ref{tab:sgrb-detection}. For optimistic cases, we can highly expect the coincident neutrino detection with GWs even with the current facilities. For the moderate case, the neutrino detection is probable with IceCube-Gen2, while it is challenging with IceCube.

\begin{table}
\centering
\caption{The detection probabilities within a given time interval $\Delta T=10$ years. The parameters used are ($\Gamma_0$, $\sigma_\Gamma)=($30, 2), (30, 4), (10, 2), and (10, 4) for EE-mod-dist-A, EE-mod-dist-B, EE-opt-dist-A, EE-opt-dist-B, respectively. This table is adapted from \cite{KMM17b}.   }
\label{tab:sgrb-detection}       
\begin{tabular}{lll}
\hline
NS-NS ($\Delta T=10$ yr)  & IC (all) & Gen2 (all) \\
\hline
EE-mod-dist-A & 0.11 -- 0.25 & 0.37 -- 0.69 \\
EE-mod-dist-B & 0.16 -- 0.35 & 0.44 -- 0.77\\
EE-opt-dist-A & 0.76 -- 0.97 & 0.98 -- 1.00 \\
EE-opt-dist-B & 0.65 -- 0.93 & 0.93 -- 1.00 \\
\hline
\end{tabular}
\end{table}

\subsection{Trans-ejecta neutrinos}\label{sec:trans}
 
The optical/UV/IR counterparts of GW170817 confirmed that neutron star mergers produce massive ejecta. The ejecta is produced immediately after the merger, while the launch of the relativistic jets can be delayed with a time lag of $t_{\rm lag}\lesssim$ 1 sec. Thus, the jets interact with the ejecta, forming a cocoon surrounding the jets \cite{RCR02a,ZWM03a,MI13b,HGN18a}. If the luminosity of the jets is low or the duration of the jet launch is short, the jets are choked inside the ejecta. This choked jet system is expected for a wide parameter range \cite{MK18a}. In this system, the photons are completely absorbed by the ejecta, while the neutrinos can penetrate the ejecta. Using such trans-ejecta neutrinos, we can discuss the physical conditions of the choked jet system without the electromagnetic signals.

\begin{figure}
\centering
\includegraphics[width=\linewidth]{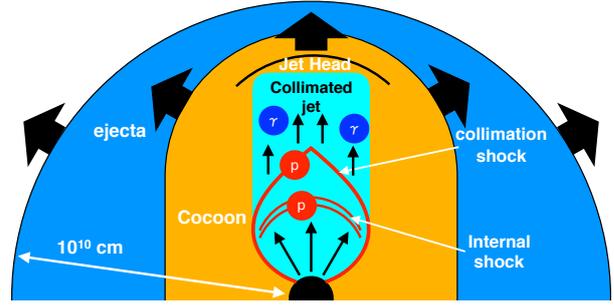}
\caption{Schematic picture of the choked jet system. This figure is reproduced from \cite{KMB18a}. }
\label{fig:trans-schematic}       
\end{figure}

In the choked jet systems, there are two possible dissipation site: the internal shocks and collimation shocks \cite{mi13}. During the jet interacting with the ejecta, the cocoon collimates the jets by pushing the jets inward, which forms the collimation shock. If the central engine creates the strong velocity fluctuation of the jets, the internal shocks can be formed below the collimation shock \cite{RM94a}. We draw a schematic picture of this system in Figure \ref{fig:trans-schematic}. The jet head, the interaction region of the jet and ejecta, also has strong forward and reverse shocks. However, we cannot expect particle acceleration there, because the density of the shock upstream is too high (see the next paragraph).

The typical size of the choked jet system is $\sim10^{10}$ cm due to the upper limit of the time lag, $t_{\rm lag}\lesssim 1$ sec. This is much smaller than the typical emission region of SGRBs (see Table \ref{tab:sgrb-param}). Hence, the dissipation region is very dense.  If the shock upstream is too dense, the shock is mediated by radiation, causing a gradual velocity change \cite{BKS10a}. This prevents the particles from being accelerated. The necessary condition for particle acceleration at the shock is given by \cite{NS12a}
\begin{equation}
 \tau_u = n_u \sigma_T l_u \lesssim 1,\label{eq:tau}
\end{equation}
where $\tau_u$ is the optical depth for the upstream, $n_u$ is the density at the upstream, $\sigma_T$ is the Thomson cross sectin, and $l_u$ is the length of the upstream fluid. To satisfy this condition, the Lorentz factor of the jets should be $\Gamma_j \gtrsim 200$ for the internal shocks and $\Gamma_j \gtrsim 500$ for the collimation shocks with a typical parameter set of the system. We set the ejecta mass $M_{\rm ej}= 0.01~\msun$, the ejecta velocity $V_{\rm ej}= 0.33c$, the lag time $t_{\rm lag}= 1$ sec, the jet opening angle $\theta_j= 0.3$ rad, the duration of the jet launch $t_{\rm dur}=2$ sec, and the kinetic luminosity $L_{k,\rm iso}= 10^{51} \rm~erg~s^{-1}$.

The neutrinos are produced at the shock downstream. The downstream of the collimation shock has a Lorentz factor of a few, leading to the very high baryon density and strong magnetic field there. This causes the strong cooling of pions, both by synchrotron and pion-proton inelastic collisions. The cutoff energy in the neutrino spectrum is typically less than 0.3 TeV, which is a too low energy to be detected by IceCube. On the other hand, the downstream of the internal shock has a relatively high Lorentz factor, $\Gamma\sim 300$, so they can emit high-energy neutrinos of $\gtrsim 100$ TeV. Thus,  we focus on the neutrinos from the internal shocks when we discuss the neutrino detectability.

\begin{figure}
\centering
\includegraphics[width=\linewidth]{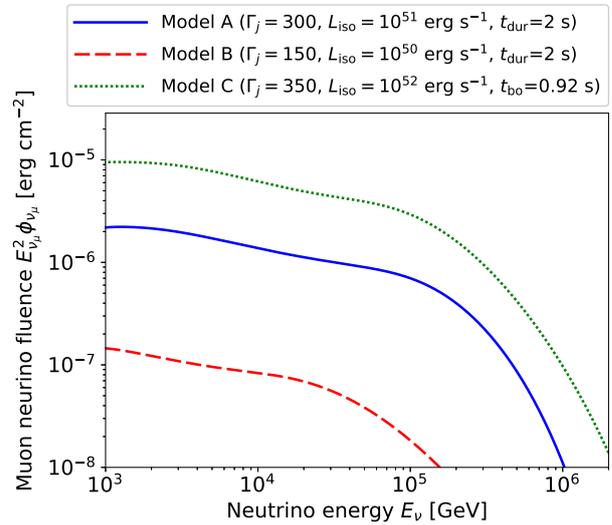}
\caption{Neutrino fluences from the internal shocks in the choked jet system. The solid and dashed lines show the optimistic and moderate cases, respectively. The dotted line indicates the precursor neutrinos of the successful jet.  This figure is reproduced from \cite{KMB18a}.  }
\label{fig:trans-fluence}       
\end{figure}

Figure \ref{fig:trans-fluence} shows the neutrino fluences from the internal shocks of the choked jet systems for the optimistic (model A) and moderate (model B) cases, whose parameters are tabulated in Table \ref{tab:trans-param}. The target photons are provided from the downstream of the collimation shock, where the photon distribution is the Planck function owing to the high optical depth. The photon density at the downstream of the internal shocks is so high that this system can be calorimetric. This leads to a flat neutrino spectrum below the cutoff energy caused by the pion cooling around 100 TeV, although the muon cooling causes a slightly softer neutrino spectrum than that for protons.  The optimistic and moderate cases differ in the Lorentz factor and the jet luminosity, which mainly change the cutoff energy and normalization of the fluence, respectively. We also plot the neutrino spectrum for a case with the successful jet case (model C). The trans-ejecta neutrinos can be produced when the jet head is inside the ejecta even for the successful jet case. Such trans-ejecta neutrinos can be detected as precursor neutrinos of the SGRBs.

Using the fluences shown in Figure \ref{fig:trans-fluence}, we estimate the detection probability of neutrinos coincident with the GWs. The upper two parts of Table \ref{tab:trans-detect} give the detection probability for a single merger event at a given distance. If the merger happens at 40 Mpc, the neutrinos are detectable with IceCube  for the optimistic case, and IceCube-Gen2 is likely to detect the neutrinos even for the moderate case. On the other hand, if the merger happens at 300 Mpc, the coincident detection is challenging with IceCube-Gen2 even for the optimistic case. The lower part provides the neutrino detection rate per year. Here, we use a neutron star merger rate obtained by the LIGO/Virgo collaborations, $\sim 1.5\times10^3\rm~Gpc^{-3}~yr^{-1}$, and consider the uniformly distributed population in the local universe. For the optimistic case, we can highly expect the coincident detection of GWs and neutrinos by IceCube with a few years of operation. Even with the moderate case, the coincident detection is likely for several years of operation with IceCube-Gen2.

\begin{table*}
\centering
\caption{The used parameters for the internal shock models. $L_{k,\rm iso}$ is the isotropic equivalent kinetic luminosity of the jets, $\Gamma_j$ is the Lorentz factor of the downstream of the internal shocks, $t_{\rm dur}$ or $t_{\rm bo}$is the duration of the jet launch (for models A and B) or the breakout time from the ejecta (for model C), $\xi_{\rm acc}$ is the baryon loading factor, and $\Gamma_{\rm rel-is}$ is the Lorentz factor of the internal shock.  This table is adapted from \cite{KMB18a}.   }
\label{tab:trans-param}       
\begin{tabular}{llllll}
\hline
model  & $L_{k,\rm iso} [\rm erg~s^{-1}]$ & $\Gamma_j$& $t_{\rm dur}$ or $t_{\rm bo}$ [s] & $\xi_{\rm acc}$ & $\Gamma_{\rm rel,is}$ \\
\hline
A & 10$^{51}$ & 300 & 2 & 1 & 4\\
B & 10$^{50}$ & 150 & 2 & 1 & 4\\
C & 10$^{52}$ & 350 & 0.92 & 1 & 4\\
\hline
\end{tabular}
\end{table*}

\begin{table*}
\centering
\caption{Detection probability of neutrinos coincident with GWs by IceCube and IceCube-Gen2 (Gen2). ``up+hor'' means the up-going and horizontal events. ``down'' means the down-going events. Since the effective area of the down-going events with IceCube-Gen2 is very uncertain, we avoid discussing it. This table is adapted from \cite{KMB18a}.}
\label{tab:trans-detect}       
Number of expected neutrinos from single event at 40\,Mpc
\begin{tabular}{llll}
\hline
model & IceCube (up+hor) & IceCube (down) & Gen2 (up+hor)\\
\hline
A & 2.0  & 0.16 & 8.7  \\
B & 0.11 & 7.0$\times10^{-3}$ & 0.46 \\
\hline
\end{tabular}
 \vspace{3pt}
 
Number of expected neutrinos from single event at 300\,Mpc
\begin{tabular}{llll}
\hline
model & IceCube (up+hor) & IceCube (down) & Gen2 (up+hor)\\
\hline
A & 0.035 & 2.9$\times10^{-3}$ & 0.15\\
B & 1.9$\times10^{-3}$ & 1.3$\times10^{-4}$ &8.1$\times10^{-3}$\\
\hline
\end{tabular}

 \vspace{5pt}
Neutrino detection rate coincident with GWs [yr$^{-1}$] 

\begin{tabular}{lll}
\hline
model & IceCube (up+hor+down)  & Gen2 (up+hor) \\
\hline
A & 0.38 & 1.2 \\
B & 0.024 & 0.091 \\
\hline
\end{tabular}
\end{table*}

\subsection{Implications from GW170817}

Although our models predict detectable neutrino fluences for some optimistic cases, such high-energy neutrinos are not detected from GW170817. However, our models are not constrained by this result. First, GW170817 turned out to be the off-axis SGRBs. For the off-axis events, the neutrino fluence decreases with $(\theta_v/\theta_j)^2$ for $\theta_v<2\theta_j $ and $(\theta_v/\theta_j)^3$ for $\theta_v>2\theta_j $ \cite{IN18a,YIN02a}, where $\theta_j$ and $\theta_v$ are the jet opening angle and the viewing angle. From the VLBI observation, these angles are estimated to be $\theta_v\sim20$ degree and $\theta_j\lesssim 5$ degree, leading to $L_v/L_{\rm on}\lesssim 1/32$, where $L_v$ and $L_{\rm on}$ are the isotropic equivalent luminosities from off- and on-axis observers, respectively. Thus, the neutrino fluence is too low to be detected by the current facilities. Second, GW170817 occurred at the southern hemisphere, where the sensitivity of IceCube is lower at $E_\nu\lesssim1$ PeV \cite{IceCube17b}. This makes it difficult to detect neutrinos of 100 TeV emitted from the choked jets and extended emissions for the optimistic model. KM3NeT will be useful to detect such neutrinos in the southern hemisphere.    Finally, GW170817 may not have the efficient neutrino production sites. The extended emission is not detected from the event, and the relativistic jet is observed from this event. Hence, it is possible to have neither the extended emissions nor the choked jets. Future on-axis events will provide detection of neutrinos or put strong constraints on the physical parameters of the choked jets and late-time activities.

\section{Super-knee cosmic rays from neutron star merger remnants}\label{sec:NSMRs}

Neutron star mergers produce fast and massive ejecta. This ejecta interacts with the ambient medium, forming a forward shock that accelerates the cosmic rays. The UV/optical/IR counterparts of GW170817 provide the mass ($0.01\rm~M_{\odot}-0.05\rm~M_{\odot}$) and velocity ($0.1c-0.3c$) of the ejecta of BNS mergers. Also, the GW observation gives rough estimate of the event rate, $\sim 1.5\times10^3\rm~Gpc^{-3}~yr^{-1}$.  This enables us to estimate the cosmic-ray production at the neutron star merger remnants (NSMRs). Since the ejecta of NSMRs are faster than that of supernova remnants (SNRs), NSMRs can produce higher energy cosmic rays than SNRs. In this section, we discuss whether the NSMRs can account for the cosmic-rays above the knee. See \cite{KMM18a} for details of this section.

At a NSMR, the balance between acceleration and age gives the maximum energy of cosmic-rays at a given time, which is expressed as 
\begin{equation}
  E_{i,\rm max} \approx \frac{3Z_i e B R_{\rm ej} V_{\rm ej}}{20 c },\label{eq:emax}
\end{equation}
where $Z_i$ is the charge of the particle species $i$, $e$ is the elementary charge, $B$ is the magnetic field, $R_{\rm ej}$ is the ejecta radius, and $V_{\rm ej}$ is the ejecta velocity. The time evolution of the velocity and radius of the ejecta is ballistic before the deceleration time, and given by the Sedov-Taylor solution after that. Then, the maximum energy of cosmic rays through an entire life of a NSMR is obtained at the deceleration time, which is estimated to be 
\begin{equation}
E_{i,\rm max} \simeq 1.8\times10^{16}Z_i \rm~eV,
\end{equation}
where we use a typical parameter set with the ejecta mass $M_{\rm ej}\simeq 0.03~\msun$, the initial ejecta velocity $V_{\rm ini}\simeq 0.25c$, the ambient density $n_{\rm amb}\simeq 0.1\rm~cm^{-3}$, and the magnetic field $B\simeq 0.4$ mG at the deceleration time.

From the GW observation, the local BNS merger rate is estimated to be $1.5\times 10^{-6}\rm~Mpc^{-3}~yr^{-1}$. Using the density of the Milky-way-size galaxies, $\sim 0.01\rm~Mpc^{-3}~yr^{-1}$, the occurrence   time of neutron star mergers in our Galaxy is estimated to be $T_{\rm mer}\simeq 1.5\times 10^{-4}~\rm yr^{-1}$. On the other hand, the escape time of the Galactic cosmic rays are estimated to be $T_{\rm esc}\sim 20-400$ Myr for particles of the rigidity $\mathcal R=10$ GV \cite{SMP07a}. The rigidity dependence of the escape time is often assumed to be $T_{\rm esc}\propto \mathcal R^{-\delta}$. For $\delta\lesssim 0.4$, the escape time for the cosmic rays of  $\mathcal R<10^8$ GV  is always longer than the occurrence time. Hence, we can use the steady state assumption. Using the grammage obtained from the recent experiments \cite{AMS16a} and one-zone approximation for the interstellar cosmic-ray density, the cosmic-ray intensity on Earth is estimated to be
\begin{equation}
 (E^2 \Phi)_i\approx \frac{(EQ_{E,\rm inj})_i X_{\rm esc}}{4\pi M_{\rm gas}}\propto E^{-\delta}\exp\left(-\frac{E}{E_{i,\rm max}}\right),
\end{equation}
where $X_{\rm esc}$ is the grammage, $EQ_{E,\rm inj}$ is the injection term, $M_{\rm gas}$ is the total gas mass in the cosmic-ray halo of the Milky-way Galaxy, $\delta\simeq1/3$ is the energy dependence of $X_{\rm esc}$, $X_{\rm esc}\propto E^{-\delta}$, for the range of our interest.
To calculate the injection term, we take into account the escape process from the NSMRs according to \cite{OMY10a}, where only the cosmic rays near the maximum energy can escape from the NSMR. Considering the time evolution of the ejecta radius and velocity discussed in the previous paragraph, the resulting escape spectrum has the same power-law index as that for the injection spectrum. The normalization of the injection term is given by $\mathcal E_{\rm cr}\approx \epsilon_{\rm cr} M_{\rm ej}V_{\rm ini}^2/2$, where we set the cosmic-ray production efficiency $\epsilon_{\rm cr}=0.25$. The composition ratio of the injection term is given by the model proposed by \cite{CYS17a}, in which the cosmic-ray injection efficiency is proportional to $(A_i/Z_i)^2$. Applying this model to the ambient medium of the solar abundance ratio, the abundance ratio of each element at the source is given by $(f_p,~ f_{\rm He},~ f_{\rm C},~ f_{\rm O},~ f_{\rm Ne},~ f_{\rm Si},~ f_{\rm Fe}) \simeq (0.17,~0.52,~0.024,~0.099,~0.027,~0.028,~0.14)$. 

\begin{figure}
\centering
\includegraphics[width=\linewidth]{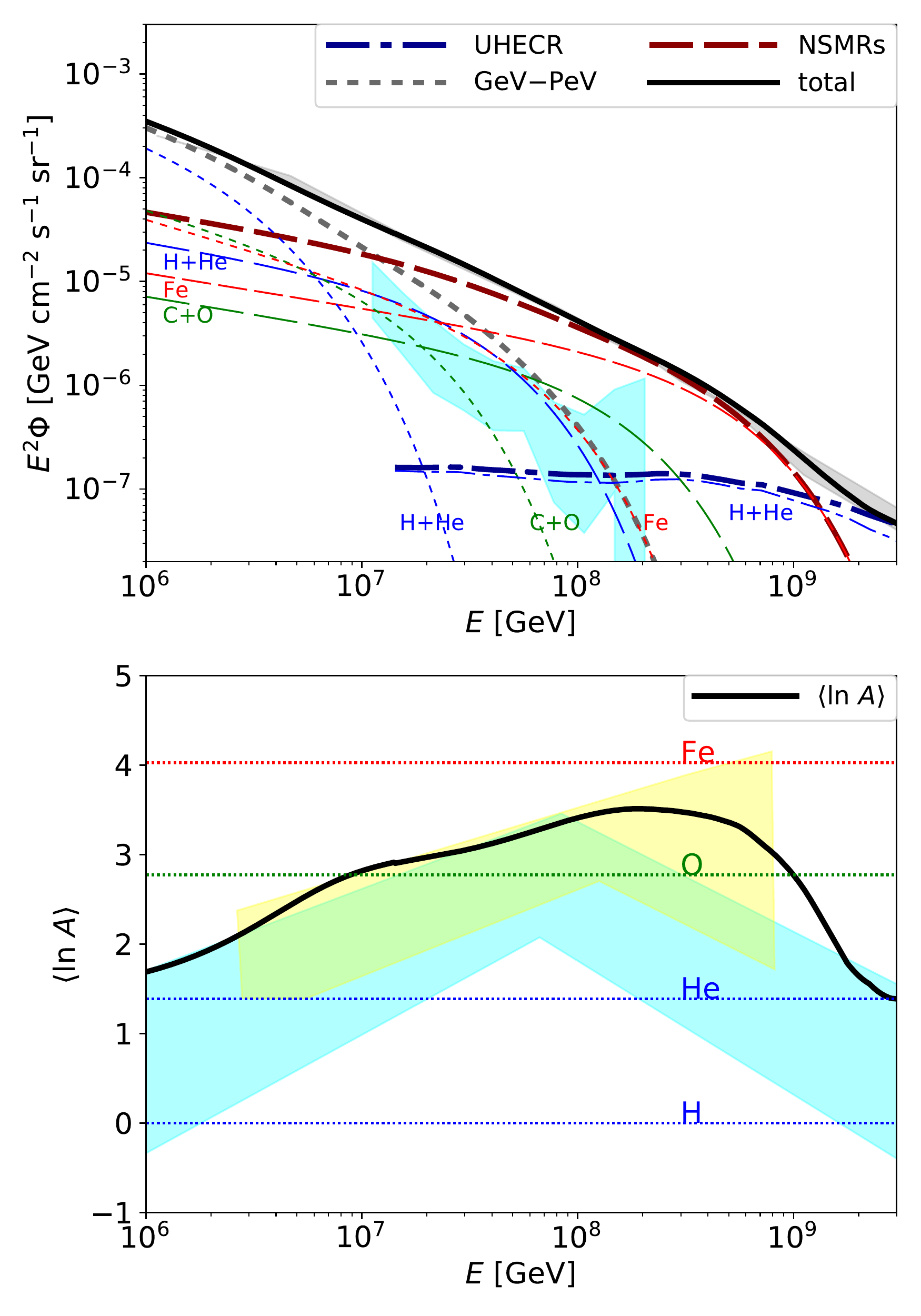}
\caption{Upper panel: Cosmic-ray intensity from NSMRs (dashed line). We also plot the GeV-PeV component (dotted line) and extragalactic component (dot-dashed line: obtained from \cite{FM18a}). The gray band represents the experimental data \cite{VIT17a,TA18a}. The cyan region shows the intensity of the light composition \cite{KASCADE13a}. Lower panel: the mean atomic number of the cosmic rays on Earth. The solid line shows our model prediction. This is consistent with the experimental results (yellow \cite{Gai16a} and cyan regions \cite{KU12a}), although uncertainty is large. This figure is reproduced from \cite{KMM18a}.}
\label{fig:GalCR}       
\end{figure}

The resulting cosmic-ray intensity is shown in Figure \ref{fig:GalCR}. The abundance ratio on Earth is written as $(f_p,~ f_{\rm He},~ f_{\rm C},~ f_{\rm O},~ f_{\rm Ne},~ f_{\rm Si},~ f_{\rm Fe}) \simeq (0.10,~0.41,~0.028,~0.13,~0.037,~0.043,~0.26)$, which is different from that at the NSMRs due to propagation effect. The GeV-PeV and extragalactic components (see \cite{FM18a} for the extragalactic one) are also plotted in the figure. Our model can reproduce the observed flux well. The cosmic-rays from NSMRs are dominant for $10^7$ GeV $\lesssim E_p\lesssim10^9$ GeV. Our model is also consistent with the hardening around $10^7$ GeV recently reported by Icetop and Telescope Array Low-energy Extension (TALE) \cite{IceTop13a,TA18a}. In addition, the light-component intensity for $10^7$ GeV $\lesssim E_p \lesssim3\times10^8$ GeV  matches that reported by KASCADE-Grande \cite{KASCADE13a}. The predicted composition ratio is also consistent with that obtained by the experiments, although the uncertainty is large.

\section{Summary}\label{sec:summary}

We have briefly reviewed the multi-messenger event GW170817, future prospects for neutrino detections coincident with GWs, and super-knee cosmic rays from the remnants of neutron star mergers. The multi-messenger campaign of GW170817 confirmed that the BNS mergers are the progenitor of SGRBs caused by relativistic jets and macronovae powered by radioactivity of r-process elements. 

The observed SGRBs are accompanied by the late-time emissions, which can emit neutrinos more efficiently than the prompt emissions. Also, if the jet duration is short or the jet luminosity is low, the ejecta of macronovae can choke the jets, causing the failed SGRBs. Such choked jets are also a strong neutrino source. Hence, the GWs can be accompanied by the high-energy neutrinos.  We estimated detectability of neutrinos from these systems within the GW horizon of the design sensitivity of advanced LIGO. For both cases, the neutrino detection coincident with the GW is probable with IceCube-Gen2 with 10-year operation even with the moderate parameter set. IceCube is also possible to detect the neutrinos with 10 years of operation for the optimistic cases.

The neutron star mergers produce massive outflows, whose velocity is higher than that of the supernovae. Thus, the NSMRs can produce higher energy cosmic rays. We estimate the cosmic-ray spectrum from Galactic NSMRs, and find that NSMRs can be the dominant source of the cosmic rays from 10 PeV to 1 EeV. Our model can naturally explain the hardening feature around 10 PeV, and the spectrum of the light elements reported by KASCADE-Grande.

\begin{acknowledgement}
S.S.K. thanks Kohta Murase and Peter Meszaros for fruitful discussion and continuous encouragement. This work is supported by JSPS oversea research fellowship and IGC fellowship.
\end{acknowledgement}

%
 \bibliography{ssk}

\end{document}